\documentclass[aps,preprint,amsmath,amssymb,amsfonts,showpacs]{revtex4}
\usepackage{epsfig}
\usepackage{graphicx}
\usepackage{subfigure}
\usepackage{dcolumn}
\usepackage{bm}

\newcommand{\const}{\mbox{const}}
\newcommand{\del}{\partial}

\renewcommand{\Re}{\operatorname{Re}}
\renewcommand{\Im}{\operatorname{Im}}
\newcommand{\sign}{\operatorname{sign}}

\newcommand{\calL}{\mathcal{L}}

\begin{document}

\title{The imaginary part of the gravitational action at asymptotic boundaries and horizons}

\author{Yasha Neiman}
\email{yashula@gmail.com}
\affiliation{Institute for Gravitation \& the Cosmos and Physics Department, Penn State, University Park, PA 16802, USA}

\date{\today}

\begin{abstract}
We study the imaginary part of the Lorentzian gravitational action for bounded regions, as described in arXiv:1301.7041. By comparing to a Euclidean calculation, we explain the agreement between the formula for this imaginary part and the formula for black hole entropy. We also clarify the topological structure of the imaginary part in Lovelock gravity. We then evaluate the action's imaginary part for some special regions. These include cylindrical slabs spanning the exterior of a stationary black hole spacetime, ``maximal diamonds'' in various symmetric spacetimes, as well as local near-horizon regions. In the first setup, the black hole's entropy and conserved charges contribute to the action's imaginary and real parts, respectively. In the other two setups, the imaginary part coincides with the relevant entropy.
\end{abstract}

\pacs{04.20.Fy,04.20.Gz,04.70.Dy}

\maketitle
\tableofcontents
\newpage

\section{Introduction} \label{sec:intro}

The action is the most important quantity in classical physics. General Relativity (GR) is the most important classical field theory. Understanding the action of GR is therefore of the utmost interest. This is not diminished by the expectation that classical GR is just a limit of some quantum theory. Indeed, whatever the underlying theory, we know that GR defines its effective large-distance action. As such, the GR action is the \emph{result} of a quantum-gravitational path integral. Thus, special properties of the GR action may tell us something about this path integral. Such an approach has been successful in the Euclidean derivation \cite{Gibbons:1976ue} of black hole entropy \cite{Bekenstein:1972tm,Bekenstein:1973ur,Hawking:1974rv}. There, the Bekenstein-Hawking entropy was extracted from an imaginary Euclidean action.

Recently, it was noticed \cite{Neiman:2012fx,Neiman:2013ap} that the gravitational action $S$ in a bounded \emph{Lorentzian} region generically has an imaginary part. This imaginary part arises from the action's boundary term, when the latter is integrated over ``flip surfaces'' - codimension-2 surfaces where the boundary flips its signature and becomes momentarily null. Such flip surfaces are always present for a closed boundary, though they may be ``hidden'' at topological corners. As the normal becomes null at a flip surface, the boundary's extrinsic curvature diverges. The resulting divergence in the action's boundary term can be resolved by a deformation of the integration contour into the complex plane. In the process, an imaginary part is picked up.

It has thus become clear that an imaginary gravitational action is a feature not just of Euclidean spacetimes, but also of finite, non-stationary regions of realistic Lorentzian solutions. Moreover, it turns out that the imaginary part of the Lorentzian action reproduces the black hole entropy formula, in the following precise way:
\begin{align}
  \Im S = \frac{1}{4}\sum_{\text{flips}}\sigma_{\text{flip}} \ . \label{eq:result}
\end{align}
Here, the sum is over all the ``flip surfaces'' in the region's boundary. $\sigma_{\mathrm{flip}}$ is the result of evaluating the black hole entropy formula on each flip surface. The relation \eqref{eq:result} was derived in \cite{Neiman:2013ap} by explicit calculation for GR with minimally-coupled matter and for Lovelock gravity. The derivation can be trivially extended to e.g. scalar fields with non-minimal couplings of the form $f(\phi)R$. For Lovelock gravity or non-minimal couplings, one should use in \eqref{eq:result} not the Bekenstein-Hawking entropy, but its generalization due to Wald \cite{Wald:1993nt,Jacobson:1993vj,Iyer:1994ys}.  

In higher-derivative theories, the entropy formula for a stationary black hole doesn't extend unambiguously to non-stationary geometries. In this case, eq. \eqref{eq:result} selects a particular non-stationary formula for $\sigma_{\mathrm{flip}}$. For Lovelock gravity, a non-trivial calculation in \cite{Neiman:2013ap} results in a formula that depends solely on the intrinsic geometry of the flip surface. This is consistent with the non-stationary entropy formula proposed in \cite{Iyer:1994ys}. 

Eq. \eqref{eq:result} cannot be applied to general diff-invariant theories, since it requires knowledge of the action's boundary term. Thus, the result's natural scope is the class of theories that contain no more than two time derivatives. In such theories, the boundary term is determined by requiring a well-defined variational principle, with only the boundary metric (along with other configuration fields) held fixed. For pure gravity, the only such theories are of the Lovelock type. Thus, the only cases not covered in \cite{Neiman:2013ap} are non-minimal matter couplings. In this paper, we will explicitly consider mostly Lovelock gravity. We do so with the understanding that adding minimally-coupled matter is trivial, and with the expectation that the same principles should apply to all two-time-derivative theories. 
 
As discussed in \cite{Neiman:2013ap}, the action's imaginary part doesn't affect the action variations that one encounters in Hamiltonian evolution. This is because such variations leave the intrinsic geometry of flip surfaces, and therefore $\Im S$, unchanged. We note here that one can also reverse that argument. Having accepted a non-vanishing $\Im S$, one can use the reality of the action variations in Hamiltonian evolution to \emph{explain} the dependence of $\Im S$ on only the intrinsic metric of the flip surfaces.

In this paper, we continue to explore the imaginary part of the gravity action. In section \ref{sec:derive}, we rederive $\Im S$ for Lovelock gravity, with GR as a special case. In the process, we draw a parallel with the Euclidean calculation in \cite{Banados:1993qp}. This enables us to explain the relation \eqref{eq:result} between $\Im S$ and the entropy formula, as well as to elucidate the topological structure of $\Im S$ in Lovelock gravity. In section \ref{sec:GH}, we focus on stationary black hole solutions, and evaluate the action for Lorentzian regions that mimic the Euclidean spacetimes of \cite{Gibbons:1976ue}. The result, given in eq. \eqref{eq:S_GH}, features a clean separation between the action's real and imaginary parts: while the real part consists of terms related to conserved charges, the imaginary part consists of terms related to the entropy. 

In section \ref{sec:diamonds}, we evaluate $\Im S$ for causal-diamond-like boundaries that span a maximal portion of some symmetric spacetime. Such boundaries coincide, in part or in full, with Killing horizons and with the spacetime's asymptotic boundary. The geometries studied include Minkowski space, the Rindler wedge, pure AdS, dS and dS$/\mathbb{Z}_2$ spaces, as well as stationary black holes with either flat or AdS asymptotics. For the regions examined, we find that $\Im S$ agrees with the relevant horizon entropy, with the exception of the de-Sitter example (there is a factor-of-2 discrepancy, which disappears if one studies dS$/\mathbb{Z}_2$ instead). In section \ref{sec:near_horizon}, we examine local near-horizon regions for both stationary and non-stationary black holes. The imaginary action for such regions again agrees with the black hole entropy. We propose a heuristic interpretation of these local regions as ``effective'' asymptotic spacetimes for highly accelerated near-horizon observers. For non-stationary black holes, the relevant horizon for this interpretation is the teleological event horizon.

It should be kept in mind that the general physical meaning of the action's imaginary part is unclear. In a Euclidean situation, it is usually understood that the imaginary action describes a partition function. In the Lorentzian, not much can be said with certainty. Considering transition amplitudes $e^{iS}$, one sees that while $\Re S$ is a phase, $\Im S$ determines the amplitude's absolute value $\left|e^{iS}\right| = e^{-\Im S}$. Thus, a positive $\Im S$ implies exponentially damped amplitudes, which one may associate with an exponentially large number $N = e^\sigma$ of available states. 

Despite the lack of a detailed physical understanding, two circumstances should be kept in mind. First, a nonvanishing $\Im S$ does come out in an honest evaluation of the boundary-term integral, and should thus be taken seriously. Second, the result \eqref{eq:result} refers primarily to finite spacetime regions, where the physical content of quantum gravity is itself unclear. Indeed, the only well-understood quantum gravity observables \cite{Witten:2001kn} are the S-matrix in asymptotically flat space and boundary CFT correlators in asymptotically AdS space \cite{Witten:1998qj,Aharony:1999ti}. In AdS/CFT, the relationship between CFT observables and bounded regions in the bulk is not entirely clear. A notable exception is the Poincare patch of AdS, which is cleanly associated with observables on a conformal Minkowski patch of the AdS boundary. As for S-matrix elements in asymptotically flat space, they are associated with the conformally compact region containing the entire spacetime. In section \ref{sec:diamonds}, we will argue that $\Im S$ actually vanishes in these two special cases. Thus, a generally non-vanishing $\Im S$ doesn't seem to contradict any existing knowledge, but may in fact further our understanding of observables in quantum gravity.

In the formulas and figures below, the spacetime metric $g_{\mu\nu}$ has mostly-plus signature. The spacetime dimension is $d\geq 2$. We use indices $(\mu,\nu,\dots)$ for spacetime coordinates, $(a,b,\dots)$ for coordinates on a codimension-1 hypersurface, and $(i,j,\dots)$ for coordinates on a codimension-2 surface. The sign convention for the Riemann tensor is $R^\mu{}_{\nu\rho\sigma}V^\nu = [\nabla_\rho,\nabla_\sigma]V^\mu$.

\section{The structure of the action's imaginary part in Lovelock gravity} \label{sec:derive}

In this section, we derive the action's imaginary part for Lovelock gravity, with GR as a special case. In sections \ref{sec:derive:intro}-\ref{sec:derive:Lorentz}, we follow the original derivation in \cite{Neiman:2013ap}, keeping track of only the essential features. Then, in section \ref{sec:derive:Euclid}, we relate the relevant piece of the Lorentzian calculation to an action in Euclidean spacetime. Using this, we'll explain in section \ref{sec:derive:entropy} why $\Im S$ agrees with the entropy formula as in \eqref{eq:result}. In section \ref{sec:derive:topology}, we will understand in topological terms why $\Im S$ depends only on the intrinsic geometry of the flip surfaces.

\subsection{Review of the Lovelock action} \label{sec:derive:intro}

\begin{figure}%
\centering%
\includegraphics[scale=0.75]{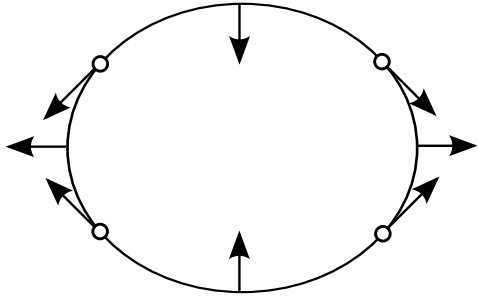} \\
\caption{A smooth closed boundary in Lorentzian spacetime. The arrows indicate the normal direction at various points. The normal's sign is chosen so that it has a positive scalar product with outgoing vectors. Empty circles denote signature flips, where the normal becomes momentarily null.}
\label{fig:smooth_normal} 
\end{figure}

The action of Lovelock gravity \cite{Lovelock:1971yv} in a spacetime region $\Omega$ consists of a bulk integral plus a boundary integral:
\begin{align}
 S = \int_\Omega \calL\, d^dx + \int_{\del\Omega} Q\, d^{d-1}x \ . \label{eq:S}
\end{align}
The Lagrangian $\calL$ is given by: 
\begin{align}
 \begin{split}
   \calL &= \sum_{m=0}^{\lfloor d/2 \rfloor} c_m \calL_m \ ; \\
   \calL_m &= \frac{(2m)!}{2^m}\sqrt{-g}\,R^{[\mu_1\nu_1}{}_{[\mu_1\nu_1}R^{\mu_2\nu_2}{}_{\mu_2\nu_2}\dots R^{\mu_m\nu_m]}{}_{\mu_m\nu_m]} \ , 
 \end{split} \label{eq:L}
\end{align}
where the $c_m$ are constant coefficients, and the upper and lower indices are antisymmetrized before tracing (it's sufficient to antisymmetrize one of the two sets). The boundary integrand $Q$ reads \cite{Myers:1987yn,Neiman:2013ap}:
\begin{align}
 \begin{split}
   Q &= \sum_{m=0}^{\lfloor d/2 \rfloor} c_m Q_m \ ; \\
   Q_m &= \sqrt{\frac{-h}{n\cdot n}}\,\sum_{p=0}^{m-1} \frac{\chi_{m,p}}{(n\cdot n)^p}
     K_{[a_1}^{[a_1} \dots K_{a_{2p+1}}^{a_{2p+1}} R^{b_1 c_1}{}_{b_1 c_1} \dots R^{b_{m-p-1} c_{m-p-1}]}{}_{b_{m-p-1} c_{m-p-1}]} \ .
 \end{split} \label{eq:Q} 
\end{align}
Here, $\chi_{m,p}$ are known numerical coefficients. $h$ is the determinant of the boundary's intrinsic metric $h_{ab}$. $n\cdot n$ is the square of the boundary normal $n^\mu$, while $K_a^b = \nabla_a n^b$ is the extrinsic curvature. Eq. \eqref{eq:Q} is written in a way that is valid for both spacelike and timelike boundaries, and is invariant under rescalings of $n^\mu$. As for the sign of $n^\mu$, it is fixed by the requirement that the \emph{covector} $n_\mu$ is outgoing, i.e. has a positive inner product with outgoing vectors. The \emph{vector} $n^\mu$ is thus outgoing, ingoing or tangent, at points where the boundary is timelike, spacelike or null, respectively. This orientation convention is summarized in figure \ref{fig:smooth_normal}.

GR with a cosmological constant is obtained as a special case, when all the $c_m$ with $m>1$ vanish. We then have:
\begin{align}
 &c_0 = \frac{\Lambda}{16\pi G};\quad \calL_0 = \sqrt{-g};\quad Q_0 = 0;
 &c_1 = \frac{1}{16\pi G};\quad \calL_1 = \sqrt{-g}R;\quad Q_1 = 2\sqrt{\frac{-h}{n\cdot n}}K \ . \label{eq:GR_coefficients}
\end{align}

In general, one can understand the Lovelock action by rearranging it as:
\begin{align}
 S = \sum_{m=0}^{\lfloor d/2 \rfloor} c_m S_m;\quad S_m = \int_\Omega\calL_m d^dx + \int_{\del\Omega} Q_m d^{d-1}x \ . \label{eq:S_m}
\end{align}
The $m$'th-order action $S_m$ is the ``dimensional extension'' of an Euler characteristic. This means that in $d=2m$ dimensions, $S_m$ is the Euler characteristic of the region $\Omega$. In $d>2m$, the integrands $\calL_m$ and $Q_m$ are given by the same expressions \eqref{eq:L}-\eqref{eq:Q} as for $d=2m$; only the dimension of the integrals in \eqref{eq:S_m} changes.

\subsection{Corner contributions}

The next step is to consider corner contributions to the action. Suppose that the boundary bends sharply by an angle $\alpha$ along some corner surface $\Sigma$. This will produce a singularity in the component $K_\bot^\bot$ of the extrinsic curvature $K_a^b$ ``along the bend''. Let us isolate a $K_\bot^\bot$ factor from $Q$ as follows:
\begin{align}
 Q \equiv \sqrt{\frac{-h}{n\cdot n}}\,K_\bot^\bot\, I + (\text{terms with no factors of } K_\bot^\bot) \ . \label{eq:isolate}
\end{align}
The corner contribution to the action, i.e. the integral of $Q$ over a small neighborhood $\delta\Sigma$ of the corner, then reads:
\begin{align}
 \int_{\delta\Sigma}Q\, d^{d-1}x = \int_\Sigma \sqrt{\gamma}\, d^{d-2}x\int d\alpha\, I \ . \label{eq:Q_corner_raw}
\end{align}
Here, $\gamma$ is the determinant of the corner's codimension-2 metric $\gamma_{ij}$, while $\alpha$ is the angle along the bend. The $d\alpha$ integral cannot be performed immediately, because, as we will see, $I$ may depend on $\alpha$.   

Due to the antisymmetrization in \eqref{eq:Q} (and ultimately, due to the two-time-derivative nature of the theory), each term in $Q$ contains at most one factor of $K_\bot^\bot$. All other indices must then be tangent to the corner surface. We denote such indices by $(i,j,\dots)$. Thus, $I$ is a polynomial in the curvature components $K_i^j$ and $R^{ij}{}_{kl}$:
\begin{align}
  I = I\left(R^{ij}{}_{kl},\, \frac{1}{n\cdot n}K_i^j K_k^l\right) \ , \label{eq:I}
\end{align}
where the second argument is invariant under rescalings of the normal. In this paper, we won't need the explicit formula for $I$. For completeness, it is given by \cite{Neiman:2013ap}:  
\begin{align}
 \begin{split}
   I ={}& \sum_{m=1}^{\lfloor d/2 \rfloor} c_m \frac{2m(2m-2)!}{2^{m-1}} \sum_{p=0}^{m-1} \binom{m-1}{p} \frac{2^{3p} (p!)^2}{(2p)!(n\cdot n)^p} \\
     &\times K_{[i_1}^{[i_1} K_{i_2}^{i_2}\dots K_{i_{2p}}^{i_{2p}} 
     R^{j_1 k_1}{}_{j_1 k_1} R^{j_2 k_2}{}_{j_2 k_2}\dots R^{j_{m-p-1} k_{m-p-1}]}{}_{j_{m-p-1} k_{m-p-1}]} \ .
 \end{split} \label{eq:I_long}
\end{align}
Naively, the extrinsic curvature components $K_i^j$ in \eqref{eq:I}-\eqref{eq:I_long} are ill-defined, since they may be discontinuous at the corner. However, it is possible to give meaning to $K_i^j$ as a function of the angle $\alpha$ along the bend.

Let us specialize to a spacelike corner $\Sigma$, which is the relevant case for signature flips. At each point of $\Sigma$, the 1+1d plane orthogonal to it is spanned by two null vectors $L^\mu$ and $\ell^\mu$. These two vectors can be viewed as null normals to the two unique lightsheets containing $\Sigma$. We can fix the scalar product $L\cdot\ell = 1$, which leaves the relative scaling of $L^\mu$ and $\ell^\mu$ arbitrary. As we traverse the corner ``along the bend'', the boundary normal $n^\mu$ gets rotated (and rescaled) in the plane spanned by $(L^\mu,\ell^\mu)$. We can thus write $n^\mu$ as a linear combination of the two null vectors:
\begin{align}
 n^\mu = \lambda(L^\mu + z\ell^\mu);\quad n\cdot n = 2\lambda^2 z \ . \label{eq:n_L_ell}
\end{align}
We imagine $z$ changing continuously from an initial value $z_1$ on one side of the corner to a final value $z_2$ on the other side. $\lambda$ may change as well, but this will have no effect on our calculation. The change in $z$ is related to the normal's boost angle as:
\begin{align}
 d\alpha = \pm \frac{dz}{2z} \ , \label{eq:alpha}
\end{align}
with the sign depending on orientation conventions. In particular, the endpoint ratio $z_2/z_1$ is related to the total corner angle $\alpha$ as $z_2/z_1 = e^{\pm 2\alpha}$. The overall scaling of $z_1$ and $z_2$ depends on the arbitrary relative scaling of $L^\mu$ and $\ell^\mu$.

Now, the extrinsic curvature components $K_i^j$ in \eqref{eq:I} can be written as:
\begin{align}
 K_i^j = \nabla_i n^j = \nabla_i(\lambda(L^j + z\ell^j)) = \lambda(\nabla_i L^j + z\nabla_i\ell^j) \ , \label{eq:K_ij}
\end{align}
where we used the fact that $L^i = \ell^i = 0$, since $L^\mu$ and $\ell^\mu$ are orthogonal to $\Sigma$. The derivatives $\nabla_i L^j$ and $\nabla_i\ell^j$ are the shear/expansion tensors of the two lightsheets generated by $L^\mu$ and $\ell^\mu$. Using eqs. \eqref{eq:I}-\eqref{eq:K_ij}, the corner contribution \eqref{eq:Q_corner_raw} takes the form:
\begin{align}
  \int_{\delta\Sigma}Q\, d^{d-1}x = \int_\Sigma \sqrt{\gamma}\, d^{d-2}x \int_{z_1}^{z_2} \frac{dz}{2z}\, 
    I\left(R^{ij}{}_{kl}, \frac{1}{2z}(\nabla_i L^j + z\nabla_i\ell^j)^2 \right) \ , \label{eq:Q_corner} 
\end{align}
where the square denotes a tensor product, with no contraction of indices. Since $I$ is polynomial in its two arguments, we see that the integrand in \eqref{eq:Q_corner} is polynomial in $R^{ij}{}_{kl}$, $\nabla_i L^j$ and $\nabla_i\ell^j$, and rational in $z$. In GR, one can read off from \eqref{eq:GR_coefficients} that $I$ is just a constant: $I = 2c_1 = 1/(8\pi G)$.  

\subsection{Extracting the imaginary part} \label{sec:derive:Lorentz}

Consider now a corner $\Sigma$ at which the boundary changes its signature e.g. from spacelike to timelike. As explained in \cite{Neiman:2013ap}, the imaginary contribution to the action at flip surfaces can always be expressed in terms of such corners. In particular, an apparently smooth signature flip (figure \ref{fig:smooth_normal}) can be thought of as a limiting case of a corner, with the corner's angle not vanishing in the limit. This is possible because a fixed spacelike-timelike corner angle can be made to appear arbitrarily ``blunt'' by boosting the reference frame. As for corners containing more than one signature flip, e.g. when the normal changes from past-pointing timelike to future-pointing timelike, their contribution to $\Im S$ is a simple multiple of that from a single flip.

Let the corner $\Sigma$, then, contain a single signature flip. In terms of the decomposition \eqref{eq:n_L_ell} of the normal $n^\mu$, this means that the coefficient $z$ changes sign, passing through zero (assuming without loss of generality that $n^\mu$ crosses through the null direction of $L^\mu$). As was noticed in \cite{SorkinThesis}, the corner angle $\Delta\alpha$ in this case has an imaginary part $\pm \pi/2$. In our present formalism, this can be seen as the result of deforming the $dz$ integral to avoid the $z=0$ pole in $d\alpha = \pm dz/2z$. The angle's imaginary part can then be read off from the closed contour integral around the pole:
\begin{align}
 \Im\int d\alpha = \pm\Im\int_{z_1}^{z_2}\frac{dz}{2z} = \pm\frac{1}{2i}\oint\frac{dz}{2z} = \pm\frac{\pi}{2} \ .
\end{align}
In GR, this result translates immediately into the action's imaginary part. In more general theories, we must perform the same contour deformation on the $dz$ integral in the corner contribution \eqref{eq:Q_corner}. The singularity at $z=0$ is no longer a simple pole, but the closed contour integral still picks out the $dz/z$ term. This leads to the following imaginary part:
\begin{align}
  \Im\int_{\delta\Sigma}Q\, d^{d-1}x = \frac{1}{2i}\int_\Sigma \sqrt{\gamma}\, d^{d-2}x \oint \frac{dz}{2z}\, 
    I\left(R^{ij}{}_{kl}, \frac{1}{2z}(\nabla_i L^j + z\nabla_i\ell^j)^2 \right) \ . \label{eq:ImQ}  
\end{align}
Summing over all flip surfaces, we express $\Im S$ as:
\begin{align}
 \Im S = \frac{1}{2i}\sum_{\text{flips}}\int_\Sigma \sqrt{\gamma}\, d^{d-2}x \oint \frac{dz}{2z}\, 
    I\left(R^{ij}{}_{kl}, \frac{1}{2z}(\nabla_i L^j + z\nabla_i\ell^j)^2 \right) \ . \label{eq:ImS}
\end{align}

The overall sign in \eqref{eq:ImQ}-\eqref{eq:ImS} depends on the orientation of the $(L^\mu,\ell^\mu)$ basis, as well as on the whether we bypass the $z=0$ singularity from above or from below. We choose the sign in such a way that (at least in GR) the action's imaginary part comes out positive. This prevents the transition amplitudes $e^{iS}\sim e^{-\Im S}$ from becoming exponentially large. We note that this sign choice can be consistently maintained only for ``causally convex'' boundaries. For arbitrary jagged boundaries, some of the flip surfaces must give negative contributions, so that the net number of flips along a closed curve in 1+1d remains four. This is necessary e.g. to ensure that the total boost angle for a closed circuit in the Lorentzian plane remains $\pm 2\pi i$. Non-convex boundaries can be accounted for in \eqref{eq:ImS} by inserting a factor of $\sign(n^\mu\del_\mu(n\cdot n))$ at each flip surface.

Corner contributions to the action with varying boundary signature have been considered before \cite{Hayward:1993my}. The recipe given there differs from ours precisely by the absence of $\Im S$. This translates into taking the principal part of the $dz$ integral \eqref{eq:Q_corner} through the $z=0$ pole. In terms of the complex $z$ plane, this principal part is the average of two proper integrals - one along a contour bypassing $z=0$ from below, and the other bypassing from above. Thus, a real action results from averaging over the two contours instead of choosing one, as we've done above. We submit that this procedure is unnatural. To reiterate the argument from \cite{Neiman:2013ap}, if both contour integrals counted as valid ``histories'', quantum mechanics would tell us to sum (or average) their two amplitudes $e^{iS}$, rather than their actions $S$. But then the contour with $\Im S < 0$ would dominate, leading to an exponentially exploding overall amplitude. We therefore believe that choosing the contour with $\Im S > 0$ is the more correct procedure.

\subsection{Relating to a Euclidean calculation} \label{sec:derive:Euclid}

In the above, we outlined the prescription \cite{Neiman:2013ap} for explicitly calculating the action's imaginary part. Our present goal is to explain the structure of the result, in particular its relation \eqref{eq:result} to the entropy formula and the topological structure of $\Im S$. Roughly speaking, we will do this by relating the closed integration contour in \eqref{eq:ImS} to a small circle in Euclidean spacetime.

Consider a closed codimension-2 surface $\Sigma$ in a Euclidean spacetime. Around each point of $\Sigma$, we draw a small circle $S_1$ in the orthogonal 2d plane. Let $D$ be the disk bounded by the $S_1$ circle. Let us now write down the gravitational action for the region $\Omega = D\times\Sigma$, with boundary $\del\Omega = S_1\times\Sigma$. As always in the Euclidean, this action will be purely imaginary. 

Since the $S_1$ circle is small, the dominant curvature in the problem is the extrinsic curvature component $K_\bot^\bot$ along the circle. Thus, the dominant contribution to the action comes from the pieces proportional to $K_\bot^\bot$ in the boundary term. We are thus led to Euclidean analogues of eqs. \eqref{eq:isolate}-\eqref{eq:I}:
\begin{align}
 Q_E &\equiv \sqrt{\frac{h}{n\cdot n}}\,K_\bot^\bot\, I + (\text{terms with no factors of } K_\bot^\bot) \ ; \\
 S_E &= -i\int_{\del\Omega}Q_E\, d^{d-1}x = -i\int_\Sigma \sqrt{\gamma}\, d^{d-2}x\int d\alpha\, 
   I\left(R^{ij}{}_{kl},\, \frac{1}{n\cdot n}K_i^j K_k^l \right) \ . \label{eq:S_euclid_raw}
\end{align}
The Euclidean factor of $-i$ can be seen as arising from the minus sign in the square root in \eqref{eq:isolate}. The overall sign follows the convention of \cite{Gibbons:1976ue}, which makes the kinetic term in $\Im S_E = -iS_E$ positive.

Now, the extrinsic curvature factors in \eqref{eq:S_euclid_raw} can again be expressed in terms of a basis of null vectors. The 2d plane of the $S_1$ circle is spanned by two \emph{complex} null vectors $(m^\mu,\bar m^\mu)$, which are complex-conjugate to each other. We again normalize so that $m\cdot\bar m = 1$, which leaves the phase of $m^\mu$ arbitrary. The boundary normal $n^\mu$ can be expressed as a linear combination of $m^\mu$ and $\bar m^\mu$ as follows:
\begin{align}
 n^\mu = \lambda(m^\mu + z\bar m^\mu);\quad n\cdot n = 2\lambda^2 z \ , \label{eq:n_m}
\end{align}
with $z$ now a complex coefficient of unit norm (the overall scaling $\lambda$ can be complex as well). As the normal is transported along the $S_1$ circle, its rotation angle is given by:
\begin{align}
 d\alpha = \pm i\frac{dz}{2z} \ . \label{eq:alpha_euclid}
\end{align}
As $n^\mu$ goes once around the $S_1$ circle, $z$ goes \emph{twice} around the unit circle in the complex plane. We are implicitly neglecting any deficit angle, since the $S_1$ circle is small. 

As in eq. \eqref{eq:K_ij}, the tangential components $K_i^j$ of the extrinsic curvature can now be expressed in terms of $\nabla_i m^j$ and $\nabla_i\bar m^j$:
\begin{align}
 K_i^j = \nabla_i n^j = \lambda(\nabla_i m^j + z\nabla_i\bar m^j) \ .
\end{align}
Plugging into the action formula \eqref{eq:S_euclid_raw}, we get:
\begin{align}
 S_E = -2\int_\Sigma \sqrt{\gamma}\, d^{d-2}x \oint \frac{dz}{2z}\, I\left(R^{ij}{}_{kl}, \frac{1}{2z}(\nabla_i L^j + z\nabla_i\ell^j)^2 \right) \ . \label{eq:S_euclid}
\end{align}

The formulas \eqref{eq:ImS} and \eqref{eq:S_euclid} are very similar. Crucially, the function $I$ in both formulas is the same. We've thus related the imaginary part of the action for an arbitrary Lorentzian region to the full action for a special Euclidean region.

\subsection{The topological structure of $\Im S$} \label{sec:derive:topology}

We can now give a topological reason for the result \cite{Neiman:2013ap} that $\Im S$ depends only on the intrinsic metric of $\Sigma$. For the Euclidean action \eqref{eq:S_euclid}, this fact was understood in \cite{Banados:1993qp}. The argument of \cite{Banados:1993qp} is as follows. Recall that up to a constant factor, the $m$'th-order term in the action \eqref{eq:S_m} is the Euler characteristic in $d=2m$ dimensions. Now, Euler characteristics are multiplicative. Thus, the $m$'th-order term in the Euclidean action \eqref{eq:S_euclid} can be decomposed into the Euler characteristic of the disk $D$ (i.e. a constant) times the Euler characteristic of $\Sigma$. But the latter can be written as a local integral using only the intrinsic metric of $\Sigma$. The same must be true for $d>2m$, since the only difference is the dimension of the integrals in \eqref{eq:S_m}.

We conclude that the Euclidean action \eqref{eq:S_euclid} can depend on $R^{ij}{}_{kl}$, $\nabla_i m^j$ and $\nabla_i\bar m^j$ only through the combination that produces the intrinsic curvature of $\Sigma$:
\begin{align}
 \tilde R^{ij}{}_{kl} = R^{ij}{}_{kl} + 4\nabla_{[k}m^{[i}\nabla_{l]}\bar m^{j]} \ .
\end{align}
Thus, we must have: 
\begin{align}
 \oint \frac{dz}{2z}\, I\left(R^{ij}{}_{kl}, \frac{1}{2z}(\nabla_i L^j + z\nabla_i\ell^j)^2 \right) = \oint \frac{dz}{2z}\, I(\tilde R^{ij}{}_{kl}, 0) 
  = i\pi I(\tilde R^{ij}{}_{kl}, 0) \ . \label{eq:I_R_tilde}
\end{align}
Note that the equality only needs to hold \emph{after} performing the $dz$ integral. The Euclidean action \eqref{eq:S_euclid} can now be rewritten as:
\begin{align}
 S_E = -2\int_\Sigma \sqrt{\gamma}\, d^{d-2}x \oint \frac{dz}{2z}\, I(\tilde R^{ij}{}_{kl}, 0)
  = -2\pi i \int_\Sigma \sqrt{\gamma}\, I(\tilde R^{ij}{}_{kl}, 0)\, d^{d-2}x \ . \label{eq:S_euclid_final}
\end{align}

Now, since the function $I$ is the same in \eqref{eq:S_euclid} and in \eqref{eq:ImS}, we conclude that the imaginary part of the Lorentzian action must also depend only on the intrinsic metric of $\Sigma$:
\begin{align}
 \Im S = \frac{1}{2i}\sum_{\text{flips}}\int_\Sigma \sqrt{\gamma}\, d^{d-2}x \oint&\frac{dz}{2z}\, I(\tilde R^{ij}{}_{kl}, 0) 
   = \frac{\pi}{2}\sum_{\text{flips}}\int_\Sigma \sqrt{\gamma}\, I(\tilde R^{ij}{}_{kl}, 0)\, d^{d-2}x \ , \label{eq:ImS_intrinsic}
\end{align}
where the codimension-2 Riemann tensor $\tilde R^{ij}{}_{kl}$ is now given by:
\begin{align}
 \tilde R^{ij}{}_{kl} = R^{ij}{}_{kl} + 4\nabla_{[k}L^{[i}\nabla_{l]}\bar\ell^{j]} \ . \label{eq:R_intrinsic}
\end{align}
The above argument complements the one in section \ref{sec:intro}, where we explained the dependence of $\Im S$ on only the flip surfaces' intrinsic metric by invoking the reality of the Hamiltonian.  

\subsection{The relation between $\Im S$ and the entropy formula} \label{sec:derive:entropy}

Another important feature of the Euclidean action \eqref{eq:S_euclid} is its direct relation to the black hole entropy formula. Given a stationary black hole spacetime, one can Wick-rotate its external region to a ``Euclidean'' section \cite{Brown:1990di}. The bifurcation surface is stable under the Wick rotation. We take it to be the surface $\Sigma$ in the ``Euclidean'' spacetime. The quotation marks denote the fact that unless the black hole is static, the Wick-rotated metric is in fact complex. Nevertheless, the analysis in section \ref{sec:derive:Euclid} carries through. In particular, on $\Sigma$ itself, and therefore approximately in its small neighborhood, the components of the Wick-rotated metric are all real. As a result, the $D\times\Sigma$ region can be constructed and analyzed just as before. Since $\Sigma$ is the image of a bifurcation surface for a pair of Killing horizons, we have $\nabla_i m^j = \nabla_i\bar m^j = 0$ and thus $\tilde R^{ij}{}_{kl} = R^{ij}{}_{kl}$ on $\Sigma$. In this case, then, the relation \eqref{eq:I_R_tilde} is trivial.

Now, the microcanonical Euclidean analysis \cite{Banados:1993qp,Brown:1995su} shows that in the above construction, the action contribution \eqref{eq:S_euclid_final} from the $S_1\times\Sigma$ boundary is just $i$ times the black hole entropy $\sigma$:
\begin{align}
 \sigma = iS_E = 2\pi\int_\Sigma \sqrt{\gamma}\, I(\tilde R^{ij}{}_{kl}, 0)\, d^{d-2}x \ . \label{eq:sigma_S_E}
\end{align}
Comparing with eq. \eqref{eq:ImS_intrinsic}, we obtain the relation \eqref{eq:result} between the imaginary part $\Im S$ of the \emph{Lorentzian} action and the black hole entropy formula.

\section{A Lorentzian version of the Euclidean black hole action} \label{sec:GH}

The importance of imaginary gravitational actions was first realized in the thermodynamic analysis of Euclidean black holes \cite{Gibbons:1976ue}, later developed in e.g. \cite{Banados:1993qp,Brown:1995su}. One goal of this paper is to clarify the relation between those Euclidean calculations and the actions of real regions in Lorentzian spacetime. So far in section \ref{sec:derive}, we related the Lorentzian imaginary action contribution from a flip surface $\Sigma$ to the action of a Euclidean region $D\times\Sigma$. We also explained the relation \eqref{eq:result} with the entropy formula by considering the special case where $\Sigma$ is the bifurcation surface of a stationary black hole. In this section, we provide a complementary picture. We will consider a stationary black hole and evaluate the action for a Lorentzian region that is analogous to the ``entire spacetime'' in the Euclidean version.

\subsection{Review of the Euclidean calculations}

First, let us recall how the action of the small region $D\times\Sigma$ arises in the Euclidean thermodynamic analysis. We will present the derivation in a way that applies to black holes in both asymptotically flat and asymptotically AdS space \cite{Hawking:1982dh,Dutta:2006vs}.

Consider, then, a stationary Lorentzian black hole. We cross over to the Euclidean by considering imaginary values of the time coordinate $t$. The imaginary time has a natural periodicity $i/T$, where $T$ is the Hawking temperature. The radial coordinate is automatically cut off from below, at the horizon radius $r = r_H$. The $t$ coordinate becomes degenerate there, so that $r = r_H$ is not a codimension-1 hypersurface but a codimension-2 surface. This is just the bifurcation surface, which we used in section \ref{sec:derive:entropy}. Thus, the Euclidean spacetime ``covers'' only the patch outside the horizon.

Consider now the action $S_E$ for the ``entire Euclidean spacetime'', i.e. for a region bounded by $r = r_\infty$ with $r_\infty$ very large. Thermal QFT suggests that $iS_E$ should be interpreted as $\ln Z$, where $Z$ is the partition function. However, $iS_E$ also contains a divergent ``baseline'', given by the action of an \emph{empty} spacetime with a given geometry of the $r = r_\infty$ slice. This baseline action can be written as $-E_0/T$, where $1/T$ is the length of the time cycle. $E_0$ depends only the intrinsic geometry of the $(r=r_\infty,\, t=\const)$ surfaces, with no dependence on the black hole's parameters. It is a divergent function of $r_\infty$. As the coefficient of the time interval in the action of empty space, we may refer to $E_0$ as a ``baseline energy''. With these preliminaries, the action of the ``entire Euclidean spacetime'' takes the form:
\begin{align}
 iS_E = \ln Z - \frac{E_0}{T} = \sigma - \frac{1}{T}(E_0 + E - \Omega J - \mu Q) \ . \label{eq:S_E_entire}
\end{align}
Here, $\Omega$, $J$, $\mu$ and $Q$ are the black hole's angular velocity, angular momentum, electric potential and charge, respectively. In higher dimensions and/or with multiple gauge fields, these quantities may carry indices, which we suppress.

Alternatively, one may consider the Euclidean black hole spacetime with the small region $D\times\Sigma$ around the bifurcation surface removed. The action \eqref{eq:S_E_entire} then acquires an extra boundary term from the new internal boundary $S_1\times\Sigma$. Another effect of removing the bifurcation surface is that the time cycle's length is no longer fixed, so we may consider actions with arbitrary values of it. The action is now composed of a bulk term and two boundary terms - one at $r_\infty$ and one at the internal boundary. Due to stationarity, these terms are all just proportional to the length of the time cycle. The proportionality constant must be the generator $-E_0 - E + \Omega J + \mu Q$ of the appropriate time translations. Setting the time cycle's length back to $1/T$, we see that the action with internal boundary must equal:
\begin{align}
 iS_E = -\frac{1}{T}(E_0 + E - \Omega J - \mu Q) \ . \label{eq:S_E_removed}
\end{align}
Comparing eqs. \eqref{eq:S_E_entire} and \eqref{eq:S_E_removed}, we conclude that the $S_1\times\Sigma$ boundary contribution, and thus the action for the $D\times\Sigma$ region, must be related to the entropy as in \eqref{eq:sigma_S_E}.

\subsection{A Lorentzian calculation}

\begin{figure}%
\centering%
\includegraphics[scale=0.75]{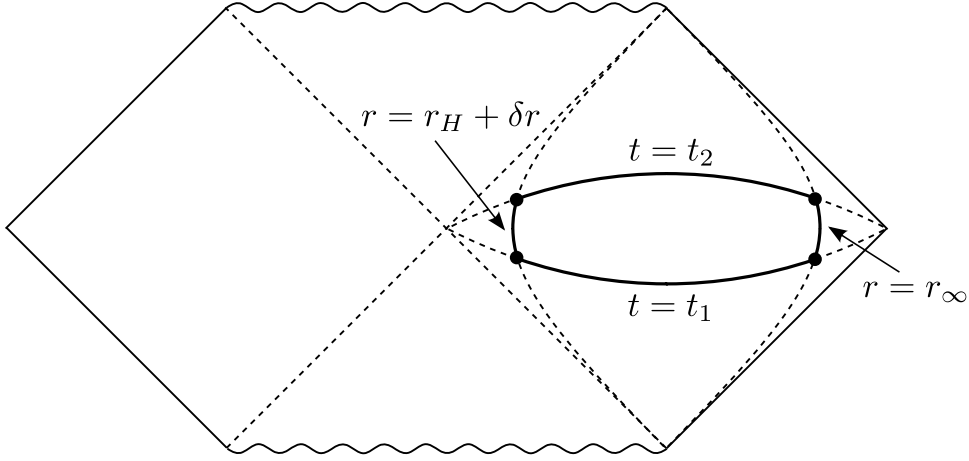} \\
\caption{A Lorentzian region in a stationary black hole spacetime, analogous to the ``entire-spacetime'' regions considered in Euclidean calculations. The Lorentzian region is bounded by two time slices and two radial slices. One radial slice is just outside the bifurcation surface, while the other is at a very large radius $r_\infty$.}
\label{fig:r_t} 
\end{figure}

We now wish to construct an analogous setup directly in the Lorentzian black hole spacetime. A complete analogy is of course impossible, since Lorentzian time is not cyclical. The closest we can get is to consider the region in the external spacetime enclosed between two time slices. Like all time slices in the black hole geometry, they intersect at the bifurcation surface and at spacelike infinity. It is helpful to ``resolve'' these two intersections by introducing two radial cutoffs, one just outside the bifurcation surface and the other at $r_\infty\rightarrow\infty$. The resulting region (for the simple example of a Schwarzschild black hole) is depicted in figure \ref{fig:r_t}. The region's boundary consists of four intersecting hypersurfaces: two constant-$t$ slices and two constant-$r$ slices. The initial and final time values $t_{1,2}$ are arbitrary. The interior radial ``cutoff'' will have no effect on the action, and only serves to make its structure more transparent. The exterior cutoff will have the same effect as in the Euclidean: it regularizes the divergent contribution associated with the ``baseline'' action of empty space. Note that regardless of the radial cutoffs, our boundary must pass near the bifurcation surface. Thus, our Lorentzian region is more closely related to the Euclidean region \emph{with} an interior boundary, whose action is given by \eqref{eq:S_E_removed}.

The action $S$ for the region in figure \ref{fig:r_t} is composed of the following terms:
\begin{enumerate}
 \item A bulk term $S_{\text{bulk}}$.
 \item A boundary term $S_H$ from the interior radial slice.
 \item A boundary term $S_\infty$ from the exterior radial slice.
 \item A boundary term $S^{(t)}_{1,2}$ from each of the time slices.
 \item A corner contribution $S^{(H)}_{1,2}$ from the intersection of each time slice with the interior radial slice.
 \item A corner contribution $S^{(\infty)}_{1,2}$ from the intersection of each time slice with the exterior radial slice.
\end{enumerate}
Note that each of the four corner surfaces contains a single signature flip.

Due to stationarity, the action's dependence on $t_1$ and $t_2$ is very simple. The only such dependence is that $S_{\text{bulk}}$, $S_H$ and $S_\infty$ are proportional to the time interval $\Delta t \equiv t_2 - t_1$. The boundary terms from the constant-$t$ slices and the corner terms do not depend on $t_{1,2}$ at all. Now, the terms proportional to $\Delta t$ are directly analogous to the corresponding terms that make up the Euclidean action \eqref{eq:S_E_removed}. By the same Noether-charge reasoning, we get:
\begin{align}
 S_{\text{bulk}} + S_H + S_\infty = -\Delta t(E_0 + E - \Omega J - \mu Q) \ . \label{eq:t_proportional}
\end{align}
As an aside, we can express the contributions $S_{\text{bulk}} + S_\infty$ and $S_H$ separately. Indeed, the $S_H$ contribution has the same form as the $S_1\times\Sigma$ Euclidean action from \eqref{eq:sigma_S_E}, with the angular interval $2\pi$ replaced by $2\pi T\Delta t$. We therefore have:
\begin{align}
 \begin{split}
   S_H &= -\Delta t\,T\sigma; \\
   S_{\text{bulk}} + S_\infty &= \Delta t(T\sigma - E_0 - E + \Omega J + \mu Q) = \Delta t(T\ln Z - E_0) \ .
 \end{split}
\end{align}

Let us now turn to the constant-$t$ boundary contributions $S^{(t)}_{1,2}$. The spacetime fields on the two time slices are all equal due to stationarity. The extrinsic curvatures, on the other hand, are equal and opposite, due to the normal's orientation. As we can see from eq. \eqref{eq:Q}, the boundary term always contains an odd power of the extrinsic curvature. Therefore, $S^{(t)}_1$ and $S^{(t)}_2$ cancel each other out (in GR, they in fact vanish separately \cite{Brown:1990di}).

To handle the corner contributions to $S$, we assume that the constant-$t$ and constant-$r$ slices are orthogonal to each other. This is essentially an assumption about the coordinate choice; it holds for e.g. Kerr-Newman black holes in GR with the standard Kerr-Newman coordinates. The orthogonality implies that the null basis vectors $(L^\mu,\ell^\mu)$ in \eqref{eq:n_L_ell} can be chosen such that $L^\mu + \ell^\mu$ is the normal to the radial slice, while $L^\mu - \ell^\mu$ is the normal to the time slice. This fixes the endpoints of the $z$ integral in the corner contribution \eqref{eq:Q_corner} to $\pm 1$.

We focus first on the corner contributions $S^{(H)}_{1,2}$ near the bifurcation surface. There, $L^\mu$ and $\ell^\mu$ are generators of the two intersecting horizons. The horizons' stationarity implies that the shear/expansion tensors $(\nabla_i L^j, \nabla_i\ell^j)$ vanish, with $R^{ij}{}_{kl} = \tilde R^{ij}{}_{kl}$ as a corollary. Substituting into eq. \eqref{eq:Q_corner}, we get:
\begin{align}
  S^{(H)}_1 = S^{(H)}_2 = \int_\Sigma \sqrt{\gamma}\, d^{d-2}x \int_{-1}^1 \frac{dz}{2z}\, I(\tilde R^{ij}{}_{kl}, 0) \ . \label{eq:S_H_raw}
\end{align}
Deforming the $dz$ integration contour as in section \ref{sec:derive:Lorentz}, we find that the integral is purely imaginary:
\begin{align}
 \int_{-1}^1 \frac{dz}{2z} = \frac{\pi i}{2} \ . \label{eq:right_angle}
\end{align}
This is just the statement that the angle between two orthogonal vectors in a Lorentzian plane is $\pi i/2$. As a result, the corner contributions $S^{(H)}_{1,2}$ are also purely imaginary. We can express them either by substituting \eqref{eq:right_angle} into \eqref{eq:S_H_raw} or by directly reading off the imaginary contribution at a flip surface from \eqref{eq:result}:
\begin{align}
 S^{(H)}_1 = S^{(H)}_2 = \frac{\pi i}{2}\int_\Sigma \sqrt{\gamma}\,I(\tilde R^{ij}{}_{kl}, 0)\, d^{d-2}x = \frac{i}{4}\,\sigma \ . \label{eq:S_H}
\end{align}
Here, $\sigma$ is the entropy functional evaluated at the bifurcation surface. Thus, it is simply the black hole's entropy.

It remains to evaluate the corner contributions $S^{(\infty)}_{1,2}$ at $r = r_\infty$. It turns out that these are also pure-imaginary, even though $\nabla_i L^j$ and $\nabla_i\ell^j$ no longer vanish. For GR, this is immediate, since the boundary term has no dependence on $(\nabla_i L^j, \nabla_i\ell^j)$ anyway. To obtain the result for higher-order Lovelock gravity, we will require one more assumption. We assume that at $r\rightarrow\infty$, the tangential extrinsic curvature components $K_i^j$ of a time slice become negligible. This can be understood in terms of the decay of higher multipoles with distance. Indeed, for a static metric, the extrinsic curvature of a time slice simply vanishes. Due to the faster decay of higher multipoles with distance, a stationary (rotating) black hole metric should tend to a static (non-rotating) one at $r\rightarrow\infty$. For example, in Kerr-Newman black holes, the $K_i^j$ components on constant-$t$ slices are smaller by a factor of $\sim 1/r^2$ than the same components on constant-$r$ slices.

Now, let us use again the orthogonality of constant-$t$ and constant-$r$ slices to choose $L^\mu - \ell^\mu$ and $L^\mu + \ell^\mu$ as their normals, respectively. Then the vanishing of $K_i^j$ on the time slice implies $\nabla_i L^j = \nabla_i\ell^j$. This makes the $I$ function in \eqref{eq:Q_corner} invariant under the substitution $z\rightarrow 1/z$, while $dz/2z$ changes sign. Thus, the substitution changes the sign of the $dz$ integral in \eqref{eq:Q_corner}. Its effect on the integration contour is to leave the endpoints $z = \pm 1$ unchanged, but flip the direction from which we bypass the $z=0$ singularity. The transformation should thus send the integral to its complex conjugate. We conclude that the corner contributions $S^{(\infty)}_{1,2}$ must equal minus their complex conjugates, i.e. they are imaginary. In analogy with \eqref{eq:S_H}, we then get:
\begin{align}
 S^{(\infty)}_1 = S^{(\infty)}_2 = \frac{i}{4}\,\sigma_0 \ ,
\end{align}
where $\sigma_0$ is the entropy functional evaluated at a $(r=r_\infty,\, t=\const)$ surface. Like $E_0$, it is a divergent function of $r_\infty$ which does not depend on the black hole's parameters. 

Putting everything together, we arrive at the following action for the Lorentzian region in figure \ref{fig:r_t}:
\begin{align}
 S = \frac{i}{2}(\sigma_0 + \sigma) - \Delta t(E_0 + E - \Omega J - \mu Q) \ . \label{eq:S_GH}
\end{align}
Here, the divergent contribution $i\sigma_0/2 - E_0\Delta t$ is the action of a cylinder in empty space with a fixed geometry of the exterior boundary at $r_\infty$. The black hole's entropy $\sigma$ contributes to the imaginary part of the action \eqref{eq:S_GH}, while the conserved charges contribute to the real part. This is in contrast with the Euclidean action \eqref{eq:S_E_entire}, where the entropy and the conserved charges contribute together.

It is tempting to derive from the action \eqref{eq:S_GH} a ``transition amplitude'':
\begin{align}
 e^{iS} = \mathcal{A}_0\, e^{-\sigma/2 - i\Delta t(E - \Omega J - \mu Q)} 
  = \mathcal{A}_0\cdot\frac{1}{\sqrt{N}}\,e^{- i\Delta t(E - \Omega J - \mu Q)} \ . \label{eq:A_GH}
\end{align}
Here, $\mathcal{A}_0\equiv e^{-\sigma_0/2 - iE_0\Delta t}$ denotes the ``amplitude of empty space'', while $N \equiv e^\sigma$ denotes the number of states associated with the black hole. Taking the square of the amplitude \eqref{eq:A_GH}, we get a ``transition probability'':
\begin{align}
 \left|e^{iS}\right|^2 = \left|\mathcal{A}_0\right|^2 \frac{1}{N} \ , \label{eq:P_GH}
\end{align} 
which appears consistent with the presence of $N$ equivalent states. The expressions \eqref{eq:A_GH}-\eqref{eq:P_GH} are suggestive, but a detailed physical picture is of course lacking. The action \eqref{eq:S_GH} may reward further study.

\begin{figure}%
\centering%
\includegraphics[scale=0.75]{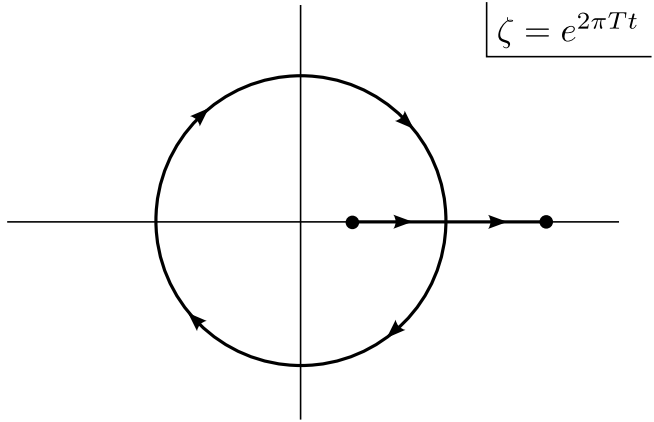} \\
\caption{Integration contours in the exponentiated time coordinate $\zeta$. The $\Delta t$-dependent piece of the action for the Lorentzian region in figure \ref{fig:r_t} is given by a line integral along the positive axis. The corresponding Euclidean action is given by a circular integral around the pole.}
\label{fig:complex_t} 
\end{figure}

Finally, with the Lorentzian action \eqref{eq:S_GH} well-defined, we can identify its precise relation to the Euclidean action with interior boundary \eqref{eq:S_E_removed}. To do this, consider the $\Delta t$-proportional terms \eqref{eq:t_proportional} in the Lorentzian action. We can rewrite the $\Delta t$ factor as a $dt$ integral; indeed, it originates from such integrals in the bulk and boundary terms. Now, near the bifurcation surface, $t$ is essentially a boost angle, in units of $2\pi T$. This implies that the coordinate that better covers the spacetime is not $t$, but $\zeta \equiv e^{2\pi Tt}$ (this coordinate of course arises in the Kruskal construction). In terms of $\zeta$, the $dt$ integral becomes:
\begin{align}
 \int dt = \frac{1}{2\pi T} \int \frac{d\zeta}{\zeta} \ . \label{eq:dt}
\end{align}
The Lorentzian action \eqref{eq:S_GH} and the Euclidean action \eqref{eq:S_E_removed} can now be treated on a similar footing. The Lorentzian action is given by a $\Delta t$-independent piece plus an integral along a real $t$ interval. The Euclidean action is given by an integral along an imaginary $t$ cycle. In the $\zeta$ plane, these two integrals translate respectively into a segment along the positive axis and a circle around the origin. See figure \ref{fig:complex_t}. Thus, the Euclidean action \eqref{eq:S_E_removed} is simply the pole residue from the integral \eqref{eq:dt}. It can therefore be viewed as the \emph{ambiguity} in the Lorentzian action due to the possibility of taking a detour around the $\zeta = 0$ pole.

\section{The imaginary part of the action for maximal diamonds} \label{sec:diamonds}

In this section, we evaluate the action's imaginary part on various causal-diamond-like regions that cover as much as possible of a given symmetric spacetime. We refer to such regions as ``maximal diamonds''. The general pattern is that $\Im S$ for such regions coincides with the entropy $\sigma$ appropriate to the situation. In most cases, the action's real part will diverge, as discussed in \cite{Neiman:2012fx}. At this point, we restrict the discussion to GR with minimally coupled matter. We will indicate the points in the argument where this restriction will play a role.

\subsection{A review of null-bounded regions} \label{sec:diamonds:null}

In figure \ref{fig:null_boundary}, we introduce some convenient terminology to describe regions with null boundaries. For such regions, the location of the flip surfaces in \eqref{eq:result} becomes ambiguous. Indeed, the boundary's normal is null on \emph{any} codimension-2 slice. As proposed in \cite{Neiman:2012fx}, the ambiguity can be resolved by considering a path integral over the different possibilities. The exponential damping $e^{iS}\sim e^{-\Im S}$ of the amplitudes picks out those slices on which $\Im S\sim\sigma_{\text{flip}}$ is minimal. On each of the lightsheets composing the boundary, one such slice will be selected and will enter the formula \eqref{eq:result}. In GR, this procedure selects the slice with the minimal area. Assuming that the area starts off decreasing from the boundary's equator towards the tips (see figure \ref{fig:null_boundary}), i.e. assuming that the equator is non-trapped, Einstein's equations ensure \cite{Bousso:1999xy} that the area will continue decreasing monotonously. Thus, the flip surfaces must be situated infinitesimally close to the tips, and eq. \eqref{eq:result} becomes:
\begin{figure}%
\centering%
\includegraphics[scale=0.75]{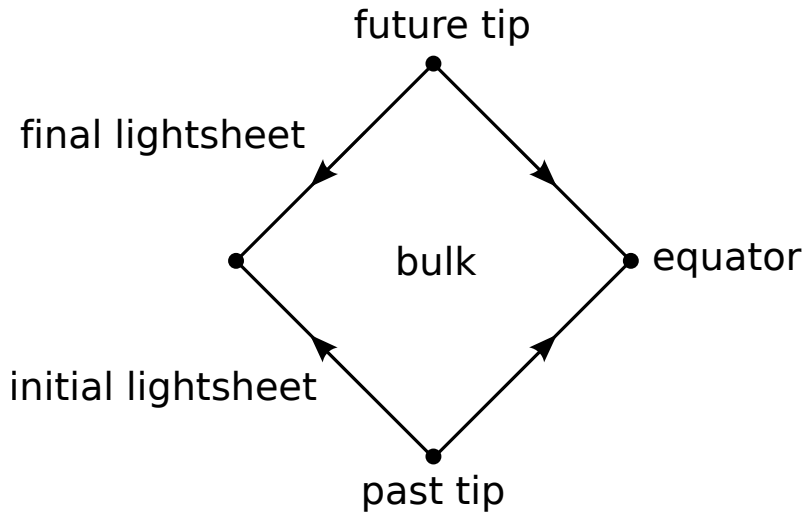} \\
\caption{Some terminology for null-bounded regions. The lines represent codimension-1 lightsheets; the dots represent codimension-2 spacelike surfaces. The arrows indicate the ``outgoing'' null normal that corresponds to the outgoing covector, for a mostly-plus metric signature. In $d>2$ dimensions with trivial topology, the left and right sides of the figure are understood to be connected.}
\label{fig:null_boundary} 
\end{figure}%
\begin{align}
 \Im S = \frac{1}{2}\sum_{\text{tips}} \sigma_{\text{tip}} = \frac{1}{8G}\sum_{\text{tips}} A_{\text{tip}} \ . \label{eq:result_null}
\end{align}
In the last equality, we used the Bekenstein entropy formula for GR. The factor of 2 between \eqref{eq:result} and \eqref{eq:result_null} arises from the two lightrays intersecting at each tip point. In other words, as we go around the boundary, the boundary's normal flips signature twice at each tip: once immediately before the tip and once immediately after. 

\begin{figure}%
\centering%
\includegraphics[scale=0.75]{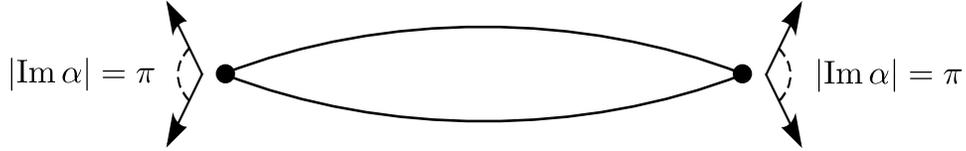} \\
\caption{A purely spacelike closed boundary, composed of two intersecting hypersurfaces. The full circles denote the intersection surface. The arrows indicate the two boundary normals at each intersection point. A continuous boost between these two normals involves two signature flips. As a result, the corner angle has an imaginary part equal to $\pi$.}
\label{fig:spacelike_boundary} 
\end{figure}

The result \eqref{eq:result_null} is unusual, in that it involves the smaller tip areas rather than the larger equator areas. As argued in \cite{Neiman:2012fx,Neiman:2013ap}, this can be understood in terms of the entropy available to an observer who starts and finishes his experiment at the tips, as opposed to the entropy that is ``potentially there in space''. Note that spacelike boundaries, as in figure \ref{fig:spacelike_boundary}, exhibit the opposite behavior. There, the signature flips lie in the boundary's ``equator-like'' corner. On the other hand, timelike boundaries behave similarly to null ones. A timelike boundary can be visualized by rotating figure \ref{fig:spacelike_boundary} on its side. Such boundaries have initial and final ``tip-like'' corners, each containing two signature flips, just like in the null case.

\subsection{Maximal diamonds} \label{sec:diamonds:maximal}

When considering boundaries at asymptotic infinity, another source of ambiguity arises: the areas of surfaces at infinity are generally ill-defined. Thus, in order to use eq. \eqref{eq:result} or \eqref{eq:result_null}, we must be clear about the limiting procedure that is employed. In each case below, we will take the limit towards asymptotic infinity in what appears to be the most natural way. Our main criterion for this ``naturalness'' is the preservation of the boundary's causal structure.

\begin{figure}%
\centering%
\includegraphics[scale=0.6]{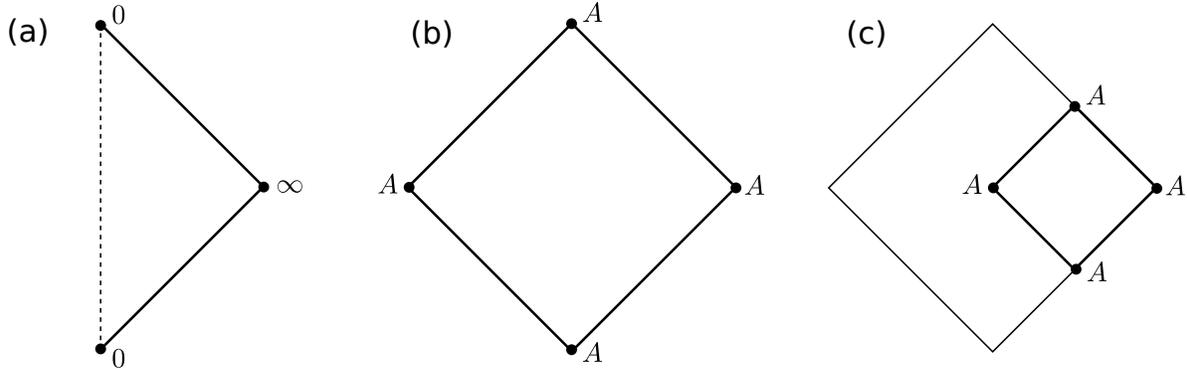} \\
\caption{Regions with null asymptotic boundaries in Minkowski-like spacetimes. (a) The region covering all of Minkowski space $\mathbb{R}^{d-1,1}$ for $d>2$. (b) The region covering an entire spacetime of the form $\mathbb{R}^{1,1}\times\mathcal{M}_{d-2}$, where $\mathcal{M}_{d-2}$ is some compact manifold with area $A$. (c) The region covering the Rindler wedge around the origin of $\mathbb{R}^{1,1}$ in the same spacetime. We have $\Im S = 0$ for case (a) and $\Im S = A/4G$ for cases (b,c).}
\label{fig:minkowski} 
\end{figure}

Consider first a region covering all of Minkowski space. In $d>2$ dimensions, the appropriate Penrose diagram is depicted in figure \ref{fig:minkowski}(a). According to the recipe in section \ref{sec:diamonds:null}, we should substitute into eq. \eqref{eq:result_null} the areas of the region's tips, i.e. the areas of past and future timelike infinity. As noted above, we will need a specific limiting procedure to make sense of this. The procedure that we propose is to view Minkowski space as the limiting case of large but finite causal diamonds, thus respecting the causal structure of conformal infinity. Now, the tip of any finite causal diamond is a single point, and its area is thus zero. In this sense, then, we should take the ``area of timelike infinity'' in Minkowski space to be zero. Plugging into eq. \eqref{eq:result_null}, we see that $\Im S$ vanishes. This is in agreement with the fact that pure Minkowski space has no entropy.

Now, consider a spacetime of the form $\mathbb{R}^{1,1}\times\mathcal{M}_{d-2}$, depicted in figure \ref{fig:minkowski}(b). Here, $\mathbb{R}^{1,1}$ is a two-dimensional Minkowski space, while $\mathcal{M}_{d-2}$ is some compact codimension-2 manifold, e.g. a torus, with area $A$. Plain two-dimensional Minkowski space can be viewed as a special case, with $A\equiv 1$. Let us find $\Im S$ for the region covering the entire $\mathbb{R}^{1,1}\times\mathcal{M}_{d-2}$ spacetime. This time, our region should be viewed as a limiting case of large finite diamonds in $\mathbb{R}^{1,1}$, each multiplied by $\mathcal{M}_{d-2}$. The tip areas for such diamonds are all $A$, so this is the area that we should use in eq. \eqref{eq:result_null}. Summing over the two tips, we find that $\Im S$ equals the entropy $\sigma = A/4G$ associated with the $\mathcal{M}_{d-2}$ slices. 

We note that a similar result is obtained for Minkowski space $\mathbb{R}^{d-1,1}$ with $d>2$ even, if the action contains a Lovelock term of order $m=d/2$. Such a term is topological, and so is its contribution to the entropy functional $\sigma_{\text{flip}}$. Therefore, as the flip surface approaches the diamond's tip, $\Im S\sim\sigma_{\text{flip}}$ will remain nonzero, even though the area vanishes. The result $\Im S = 1/4G$ for GR in two dimensions can be viewed as a special case of this topological mechanism.

In addition to considering an entire $\mathbb{R}^{1,1}\times\mathcal{M}_{d-2}$ spacetime, we can calculate $\Im S$ for a Rindler wedge that covers a quarter of $\mathbb{R}^{1,1}$. See figure \ref{fig:minkowski}(c). There is a two-parameter family of such regions, depending on the choice of origin in $\mathbb{R}^{1,1}$. Like in the full $\mathbb{R}^{1,1}\times\mathcal{M}_{d-2}$ spacetime, the tip areas are all $A$, and we again get $\Im S = A/4G$. This is the entropy that one usually associates to a Rindler wedge. In section \ref{sec:near_horizon}, we will apply these results to near-horizon regions in black hole spacetimes.

\begin{figure}%
\centering%
\includegraphics[scale=0.6]{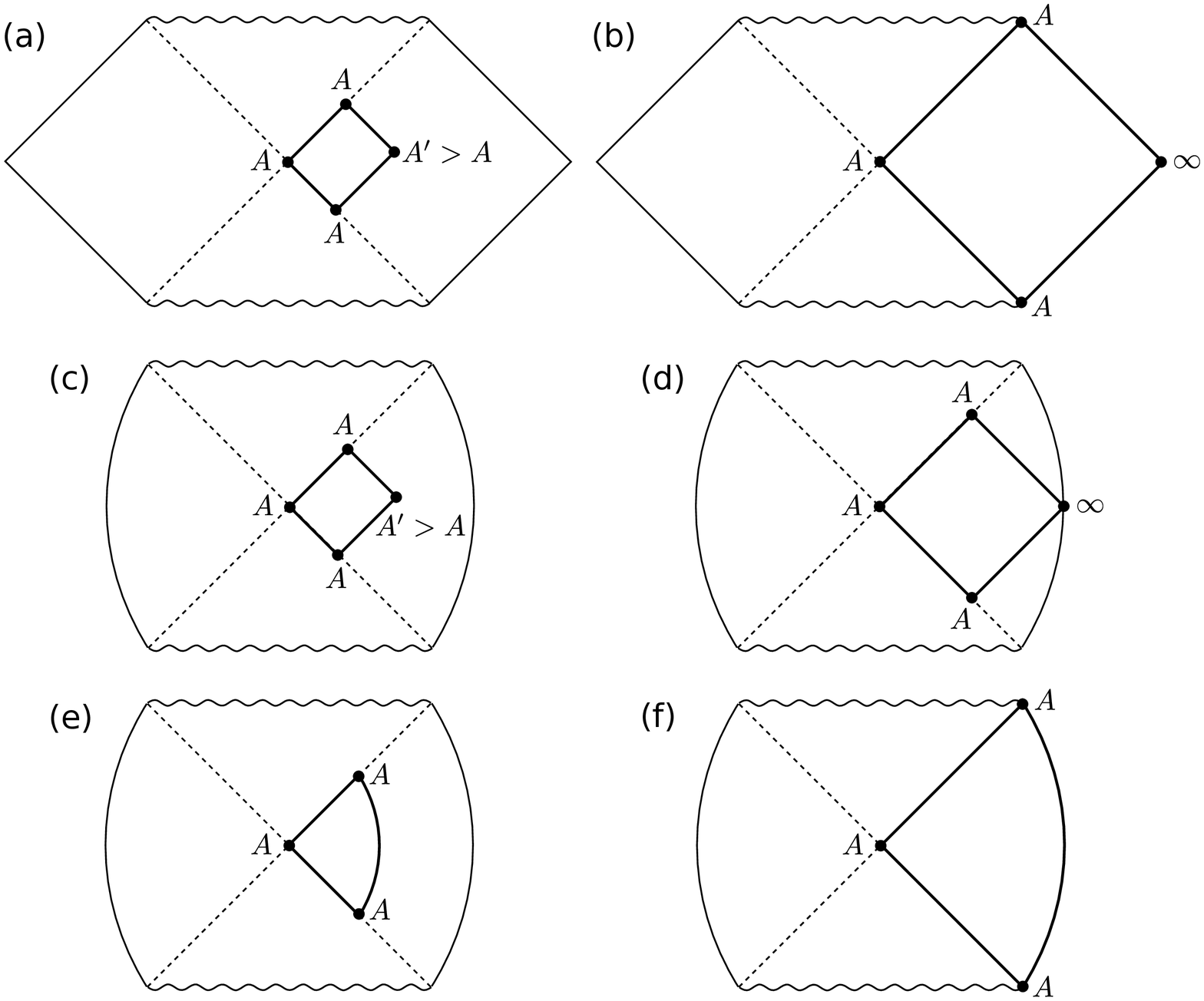} \\
\caption{Regions in the external spacetime of stationary black holes. In (a)-(b), the spacetime is asymptotically flat, while in (c)-(f) it is asymptotically AdS. Figures (a,c,e) depict finite regions, which through a limiting procedure become the asymptotic regions in (b,d,f), respectively. In every case, the action's imaginary part equals the black hole's entropy.}
\label{fig:kruskal} 
\end{figure}

We now turn to ``maximal diamonds'' in the external spacetime regions of stationary black holes. These are depicted in figure \ref{fig:kruskal}. The Penrose diagrams in the figure are for Schwarzschild black holes, but as in figure \ref{fig:r_t}, this is only for concreteness. What we require is a stationary spacetime with an exterior asymptotic region and a bifurcate Killing horizon. Figure \ref{fig:kruskal}(b) depicts the null-bounded region that covers the entire external spacetime for an asymptotically flat black hole. Figure \ref{fig:kruskal}(d) depicts a ``maximal'' null-bounded region for an asymptotically AdS black hole. This region does not cover the entire external spacetime, and only touches spatial infinity at a single codimension-2 slice. There is an infinite-dimensional family of such regions, one for each choice of this slice. Figure \ref{fig:kruskal}(f) depicts the region that \emph{does} cover the entire external spacetime in asymptotically AdS. This region's boundary is partially null (along the event horizons) and partially timelike (along the AdS boundary). Figures \ref{fig:kruskal}(a,c,e), depict the finite regions from which the asymptotic regions in figures \ref{fig:kruskal}(b,d,f) are obtained as limiting cases (the diamond in figure \ref{fig:kruskal}(a) was already considered in \cite{Neiman:2013ap}). Applying eq. \eqref{eq:result_null} to each of these finite regions, we find that $\Im S$ coincides with the black hole's entropy $\sigma$ in every case. The limiting procedure then suggests that the same is true for the asymptotic regions in figures \ref{fig:kruskal}(b,d,f).

Incidentally, the regions in figure \ref{fig:kruskal} offer some insight into the fact that $\Im S$ for null boundaries is determined by the tip areas. As discussed in section \ref{sec:diamonds:null}, this implies that with regard to $\Im S$, null boundaries behave like timelike ones. In the present context, this maps to similar behavior for the null and timelike asymptotic boundaries of asymptotically flat and AdS spacetimes, respectively. In turn, this leads to the agreement on the result $\Im S = \sigma$ between figures \ref{fig:kruskal}(a,b) and \ref{fig:kruskal}(e,f). Thus, the statement ``null boundaries are like timelike ones (as opposed to spacelike)'' becomes ``flat spacetime is like AdS (as opposed to de-Sitter)''. When discussing the action and entropy of stationary black holes, this indeed makes sense: asymptotically flat and AdS spacetimes admit such solutions, while asymptotically de-Sitter ones do not.

\begin{figure}%
\centering%
\includegraphics[scale=0.6]{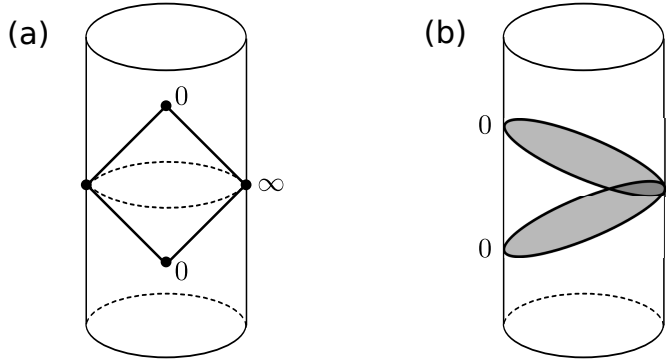} \\
\caption{Two kinds of ``maximal diamonds'' in anti-de-Sitter space: (a) a causal diamond with tip points in the bulk and an equator at spatial infinity; (b) a Poincare patch. In both cases, the action's imaginary part vanishes.}
\label{fig:AdS} 
\end{figure}

It is also interesting to consider the analogue of figure \ref{fig:minkowski}(a) in pure de-Sitter and AdS spacetimes. Unlike in Minkowski space, there is no longer a causal diamond that covers the entire spacetime. Let us focus first on pure AdS space. Here, two possible notions of a ``maximal diamond'' come to mind. The first of these is depicted in figure \ref{fig:AdS}(a). It is obtained by choosing a bulk point, following its future lightcone up to the AdS boundary, letting the rays ``bounce back'' and following them again until they refocus. There is a $d$-parameter family of such diamonds, determined by the location of the initial tip. Since these diamonds have pointlike tips with vanishing area, we get $\Im S = 0$. This is consistent with the fact that pure AdS has zero entropy. The asymptotically AdS diamonds in figures \ref{fig:AdS}(a) and \ref{fig:kruskal}(d) are related in the same way as the asymptotically flat diamonds in figures \ref{fig:minkowski}(a) and \ref{fig:kruskal}(b).

The other possible notion of a ``maximal diamond'' in AdS is a Poincare patch, depicted in figure \ref{fig:AdS}(b). The boundary of the Poincare patch consists of the timelike AdS boundary and two spacelike hypersurfaces, one initial and one final. At spatial infinity, the two spacelike boundaries become asymptotically null. There is a $(d-1)$-parameter family of Poincare patches, depending e.g. on the intersection point of the spacelike hypersurfaces at spatial infinity. The signature flips on the boundary of the Poincare patch occur at the codimension-2 intersections between its spacelike and timelike components. While these intersections are not pointlike, they are null, and should thus be associated with a vanishing area (in the conformal geometry of the AdS boundary, they play the roles of past and future null infinity). Thus, we again get $\Im S = 0$, in accordance with the vanishing entropy of a Poincare patch in AdS. Note that we used here the restriction to GR with minimal couplings: otherwise, it isn't obvious that $\Im S\sim\sigma_{\text{flip}}$ vanishes for flip surfaces with vanishing area.

\begin{figure}%
\centering%
\includegraphics[scale=0.75]{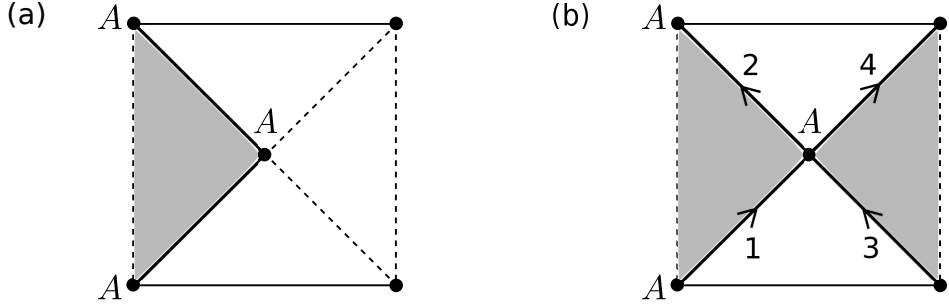} \\
\caption{``Maximal diamonds'' (shaded) in (a) standard de-Sitter space $dS_d$ and (b) elliptic de-Sitter space $dS_d/\mathbb{Z}_2$. Every point in the Penrose diagrams is a sphere $S_{d-2}$ in spacetime. In (b), each sphere is identified with the sphere that lies opposite from it across the diagram's center. Each point on one sphere is identified with the antipodal point on the other. The numbered arrows trace a continuous null path along the diamond's boundary.}
\label{fig:dS} 
\end{figure}

Finally, let us turn to de-Sitter space. There, the natural candidate for a ``maximal diamond'' is a null-bounded region whose equator is a codimension-2 equatorial sphere. Without loss of generality, we can place this sphere at the center of the standard square Penrose diagram. The two lightsheets passing through the equatorial sphere have a constant cross-section area $A$. They never focus at a finite bulk point, but instead focus conformally at timelike infinity. One can define a maximal diamond by stretching two non-parallel lightsheet segments from the equator to timelike infinity, one past-going and the other future-going. The resulting shape is depicted as the shaded region in figure \ref{fig:dS}(a). There is a $2(d-1)$-parameter family of such diamonds, determined by the positions of the two tip points.

Along each of the two lightsheets in figure \ref{fig:dS}(a), there is a spherical signature-flip surface. Each of these flip surfaces contributes $\sigma_{\text{flip}}/4 = A/(16G)$ to $\Im S$. The imaginary part of the action thus evaluates to: 
\begin{align}
 \Im S = \frac{A}{8G} \ ,
\end{align}
which is \emph{half} of the entropy $\sigma = A/4G$ usually associated with de-Sitter space (see e.g. \cite{Gibbons:1976ue}). This is a somewhat disappointing result, since it breaks the $\Im S = \sigma$ pattern.

It turns out that the mismatch disappears if we consider not a standard de-Sitter space $dS_d$, but an ``elliptical'' de-Sitter space $dS_d/\mathbb{Z}_2$. This space is obtained by identifying antipodal points related through $CPT$. It has been argued \cite{Parikh:2002py} that $dS_d/\mathbb{Z}_2$ is a more suitable arena for quantum gravity than $dS_d$. The ``maximal diamond'' in $dS_d/\mathbb{Z}_2$ is depicted as the shaded region in figure \ref{fig:dS}(b). Each point on the Penrose diagram is identified with the opposite point across the diagram's center. In particular, opposite corners of the square are identified. These identifications involve also passing to the antipodal point in the $S_{d-2}$ spheres that reside at each point of the 2d diagram. 

The arrows in figure \ref{fig:dS}(b) trace a typical continuous null path around the boundary of the ``maximal diamond''. The numbers indicate the order in which the four segments are traversed. Segments 3-4 are at an antipodal point of the $S_{d-2}$ sphere from segments 1-2. There is no global time orientation in $dS_d/\mathbb{Z}_2$, and the same is true of our ``maximal diamond''. However, the local causal structure of the diamond's boundary is almost the same as the standard one in figure \ref{fig:null_boundary}. In particular, the center of the Penrose diagram functions as an equator, while the two inequivalent corners function as tips. The only local difference is that now at every equator point, all four transverse null directions belong to the boundary. As always in $dS_d/\mathbb{Z}_2$, the overall inconsistency of the time orientation is apparent only to observers at timelike infinity.

Let us now calculate $\Im S$ for the diamond in figure \ref{fig:dS}(b). Since antipodal points are identified, we may focus on e.g. the left half of the Penrose diagram. There, we again have a single flip surface on each of the two lightsheets. Each flip surface again contributes $A/(16G)$ to the action's imaginary part. Overall, we get:
\begin{align}
 \Im S = \frac{A}{8G} = \frac{A_{1/2}}{4G} \ , \label{eq:S_dS_Z2}
\end{align}
where $A_{1/2} = A/2$ is the area of the codimension-2 equator sphere after the $\mathbb{Z}_2$ identification. Thus, the imaginary action \eqref{eq:S_dS_Z2} coincides with the entropy $\sigma_{1/2} = A_{1/2}/4G$ of the ``halved'' de-Sitter space $dS_d/\mathbb{Z}_2$.

\section{Near-horizon regions and non-stationary black holes} \label{sec:near_horizon}

In sections \ref{sec:GH}-\ref{sec:diamonds}, we evaluated $\Im S$ for a selection of regions. These regions were all situated in symmetric spacetimes and involved knowledge of asymptotic infinity. These are the contexts in which our understanding of quantum gravity and black hole entropy is at its firmest. Thus, the analysis in sections \ref{sec:GH}-\ref{sec:diamonds} is essential in order to make contact with existing approaches.

However, in some sense, this analysis misses the conceptual novelty of eq. \eqref{eq:result}. Indeed, it's important that \eqref{eq:result} not only knows about the entropy formula, but is also very general: it applies to arbitrary regions, with no requirements of symmetry or asymptotics. It may therefore provide a window into the situations where quantum gravity is conceptually hard. Thus, along with stationary asymptotic regions, it is important to also study regions that are more dynamical and local in nature.

The first item on this agenda should be to say something about non-stationary black holes. Indeed, the discussion in section \ref{sec:diamonds:maximal} (as well as the examples in \cite{Neiman:2013ap}) assumed perfect stationarity, including a bifurcate Killing horizon. This is a very restrictive assumption, which excludes all astrophysical black holes - even ones that settle into stationarity after their formation. In addition, realistic black holes exist in a universe which, by our present understanding of cosmology, asymptotes in the future to de-Sitter space. Thus, the flat or AdS asymptotics from figure \ref{fig:kruskal} are most probably wrong. In any case, accelerated-expansion cosmology renders spatial infinity inaccessible. Therefore, in addition to non-stationarity, it is important to consider \emph{local} regions in black hole spacetimes, without assuming knowledge of the asymptotics. In this section, we make tentative steps in these directions. Since we'll be considering boundaries with nearly constant cross-sections, we no longer require the area monotonicity results from GR. Thus, the discussion below applies to general theories of gravity with two time derivatives.  

\subsection{The near-horizon limit with a bifurcation surface} \label{sec:near_horizon:bifurcation}

Consider again a stationary black hole. A standard trick is to approximate the near-horizon region with a Rindler geometry. In this picture, the accelerated observers outside the horizon get reinterpreted as accelerated Rindler observers. The origin of the Rindler wedge corresponds to the black hole's bifurcation surface. It is then interesting to ask, what does the asymptotic boundary of the Rindler space correspond to? Clearly, it doesn't exist anywhere in the original black hole spacetime, since the Rindler approximation breaks down at large radii. However, we propose that it's useful to \emph{imagine} the asymptotic boundary of the Rindler space situated a small distance outside the black hole horizon. This encapsulates the notion of the Rindler space being a near-horizon approximation: the farthest distances in the Rindler space correspond to small finite distances in the black hole spacetime.   

Pictorially, we propose to associate a small finite diamond as in figure \ref{fig:kruskal}(a) with the entire Rindler wedge from figure \ref{fig:minkowski}(c). The codimension-2 section $\mathcal{M}_{d-2}$ of the Rindler wedge is then identified with the bifurcation surface. The smallness of the finite diamond is expressed through the fact that this section remains approximately constant. In particular, it implies that $A'\approx A$ in figure \ref{fig:kruskal}(a). We note that a finite diamond cannot capture the full domain of the near-horizon approximation, which is valid along the entire extent of the bifurcate horizon. However, the diamond can be made arbitrarily long in one of the null directions, while becoming narrower in the other. In this way, a single diamond can capture the near-horizon region arbitrarily far along one horizon branch, e.g. along the future horizon.    

Now, as calculated in section \ref{sec:diamonds:maximal}, the action's imaginary part for both the finite diamond and the infinite Rindler wedge reproduces the black hole's entropy $\sigma$ (due to the constant cross-section, the generalization beyond GR is straightforward). We have, then, a physical picture that involves an agreement between $\Im S$ and $\sigma$, but makes no mention of the asymptotics in the black hole spacetime.       

\subsection{Local regions near non-stationary event horizons}

The next step is to do away with the bifurcation surface. One way is to consider a Rindler wedge centered at an arbitrary horizon slice. However, this construction is not very well-motivated. Another approach is to draw a lesson from figures \ref{fig:minkowski}(b,c). There, we've seen that as far as $\Im S$ is concerned, there is no difference between a Rindler wedge and an entire $\mathbb{R}^{1,1}\times\mathcal{M}_{d-2}$ spacetime. Thus, it's not really necessary to fix an origin for a Rindler wedge. 

\begin{figure}%
\centering%
\includegraphics[scale=0.6]{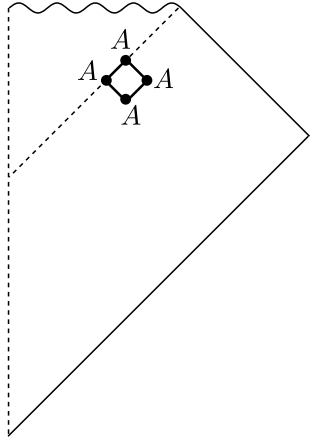} \\
\caption{A local near-horizon region in a non-eternal, non-stationary black hole spacetime. The imaginary part of this region's action reproduces the instantaneous entropy of the black hole. The relevant horizon for the discussion in the text is the event horizon.}
\label{fig:local} 
\end{figure}

Consider, then, some slice $\Sigma$ of the event horizon for some non-eternal, non-stationary black hole (figure \ref{fig:local}). An observer that approaches $\Sigma$ from large radii and then escapes to infinity must be highly accelerated. One can imagine a set of such observers spanning the horizon slice, each with a worldline that is highly accelerated in the transverse 1+1d plane. The observers' high acceleration defines a small distance scale. Therefore, one may say that they see a ``zoomed-in'' version of the surrounding spacetime. Since the observers' acceleration is higher than any curvature in the spacetime itself, this ``zoomed-in'' spacetime is just $\mathbb{R}^{1,1}\times\Sigma$, with $\mathbb{R}^{1,1}$ a flat 1+1d Minkowski space. As in section \ref{sec:near_horizon:bifurcation}, we can visualize the asymptotic boundary of $\mathbb{R}^{1,1}\times\Sigma$ as occupying a small near-horizon diamond, depicted in figure \ref{fig:local}. The precise size and location of this diamond are of course ill-defined. However, these details are irrelevant for calculating $\Im S$. We again find that for small diamonds, the value of $\Im S$ coincides with the black hole's entropy $\sigma$. Note that this $\sigma$ is the \emph{instantaneous} entropy, as evaluated on the chosen horizon slice using only its intrinsic geometry.  

The accelerated observers in the above argument do not need to be Rindler observers. This means that an observer's distance from the horizon (as measured in his reference frame) need not coincide with the inverse of his acceleration. It is in this sense that we refer to an effective $\mathbb{R}^{1,1}\times\Sigma$ spacetime rather than to a Rindler wedge. 

How much of the above picture is sensitive to asymptotic infinity? In our physical justification for the highly accelerated observers, we used the notion of ``escaping to infinity''. Therefore, the relevant horizon in the discussion is the teleological event horizon, rather than some more local structure. Of course, one could choose a different, finite, hypersurface as the criterion for ``escape''. However, such a construction would still be teleological. In contrast, the actual calculation of $\Im S$ in figure \ref{fig:local} is completely local. In fact, $\Im S$ for \emph{any} small 1+1d region times a codimension-2 surface $\Sigma$ will give the entropy functional evaluated on $\Sigma$. The challenge is to incorporate this fact into some physically compelling picture.    

\section{Discussion} \label{sec:discuss}

In this paper, we expanded the study of the imaginary part of the Lorentzian gravitational action. The surprising results of the brute-force calculations in \cite{Neiman:2013ap} were explained more abstractly, through a relation with Euclidean geometries. We also turned around the discussion in \cite{Neiman:2013ap} concerning the reality of the Hamiltonian, using it to \emph{explain} the dependence of $\Im S$ in Lovelock gravity on only the intrinsic metric of the flip surfaces.   

We then evaluated $\Im S$ for three classes of physically interesting regions. For spacelike-bounded regions describing time evolution outside a stationary black hole (section \ref{sec:GH}), we found that the black hole's entropy and conserved charges enter $\Im S$ and $\Re S$, respectively. We also found that the ``transition probability'' $\left|e^{iS}\right|^2$ is inversely proportional to the number of microstates $N = e^\sigma$ implied by the black hole's entropy. It would be interesting to see if this can be fleshed out into a more detailed physical picture.

For null-bounded ``maximal diamonds'' in various symmetric spacetimes, we found that $\Im S$ coincides with the entropy that is usually associated to each spacetime. This is an intriguing pattern that calls for a deeper understanding. In particular, it can be taken to support the notion \cite{Neiman:2012fx} that null boundaries are best suited for the study of quantum gravity. Standard de-Sitter space poses an exception to the $\Im S = \sigma$ pattern, which is resolved if one considers the ``elliptical'' de-Sitter space dS$/\mathbb{Z}_2$ instead. If one takes $\Im S$ seriously as a window into quantum gravity, this may serve as yet another argument favoring dS$/\mathbb{Z}_2$ over dS as the more appropriate spacetime asymptotics.  

Finally, we discussed regions composed of small 1+1d diamonds times codimension-2 surfaces $\Sigma$. For all such regions, $\Im S$ evaluates to the entropy as calculated from $\Sigma$. We proposed a physical interpretation for this calculation in two setups: a near-bifurcation-surface region in a stationary black hole spacetime and a near-event-horizon region for a non-stationary black hole.

In the future, it would be interesting to try and relate $\Im S$ to the concept of entanglement entropy, both in AdS/CFT \cite{Nishioka:2009un} and more generally in quantum field theory. 

On a more fundamental note, one would like to address the conceptual issues that arise from the non-unitarity of amplitudes $e^{iS}$ for complex $S$. This may serve as a concrete starting point on the broader mystery of the physical content of quantum gravity in finite regions. 

\section*{Acknowledgements}		

I am grateful to Abhay Ashtekar, Norbert Bodendorfer, Logan Ramalingam, Eugenio Bianchi and Rob Myers for discussions. This work is supported in part by the NSF grant PHY-1205388 and the Eberly Research Funds of Penn State. Parts of the work were carried out at Harvard University and Perimeter Institute.

\end{document}